\newif\ifdraft
\newif\iffull
\newif\ifcomment
\newif\iflatexdiff
\newif\ifbibtex
\newif\ifpreprint
\preprinttrue    
\fulltrue        
\def\dvers{v1.02}
\def\snntitle{$\snn$}
\ifpreprint
\def\snntitle{$\snnbf$}
\fi
\def\dtitle{Long-range angular correlations on the near and away side\\ in p--Pb collisions at \snntitle = 5.02 TeV} 
\def\stitle{Long-range angular correlations in p--Pb collisions} 
\ifpreprint
\documentclass[ALICE,manyauthors,12pt]{cernphprep}
\usepackage[comma,square,numbers,sort&compress]{natbib}
\usepackage{hyperref}
\else
\documentclass[final,3p,12pt]{elsarticle}
\biboptions{comma,square,numbers,sort&compress}
\usepackage{hyperref}
\fi
\usepackage{graphicx}  
\usepackage{dcolumn}   
\usepackage{bm}        
\usepackage{amssymb}   
\usepackage{amsfonts}
\usepackage{graphics}
\usepackage{grffile}   
\usepackage{epsfig}
\usepackage{units}
\usepackage[usenames]{color}
\usepackage[normalem]{ulem} 
\usepackage{color}
\usepackage[utf8]{inputenc}
\usepackage[T1]{fontenc}
\iflatexdiff
\RequirePackage{color}\definecolor{RED}{rgb}{1,0,0}\definecolor{BLUE}{rgb}{0,0,1}

\fi
\ifpreprint
\else
 
\fi

\newcommand{\gevc}         {GeV/\ensuremath{c}}

\newcommand{\ptt}          {\ensuremath{p_{\mathrm{T, trig}}}}
\newcommand{\pta}          {\ensuremath{p_{\mathrm{T, assoc}}}}

\newcommand{\ITS}          {\rm{ITS}}

\newcommand{\ZNA}          {\rm{ZNA}}
\newcommand{\ZNC}          {\rm{ZNC}}

\newcommand{\ZDCs}         {\rm{ZDCs}}
\newcommand{\SPD}          {\rm{SPD}}
\newcommand{\SDD}          {\rm{SDD}}

\newcommand{\TPC}          {\rm{TPC}}
\newcommand{\VZERO}        {\rm{VZERO}}
\newcommand{\VZEROA}       {\rm{VZERO-A}}
\newcommand{\VZEROC}       {\rm{VZERO-C}}

\newcommand{\pp}           {pp}

\newcommand{\pPb}          {\mbox{p--Pb}}

\newcommand{\dAu}          {\mbox{d--Au}}

\newcommand{\dNdeta}       {\mathrm{d}N_\mathrm{ch}/\mathrm{d}\eta}

\newcommand{\s}            {\ensuremath{\sqrt{s}}}
\newcommand{\pt}           {\ensuremath{p_{\mathrm{T}}}{ }}

\newcommand{\snn}          {\ensuremath{\sqrt{s_{\mathrm{NN}}}}}
\newcommand{\snnbf}        {\ensuremath{\mathbf{{\sqrt{s_{\mathbf NN}}}}}}

\newcommand{\avg}[1]       {\ensuremath{\left\langle#1\right\rangle}}

\newcommand{\dd}           {\ensuremath{\mathrm{d}}}

\newcommand{\Dphi}         {\ensuremath{\Delta\varphi}}
\newcommand{\Deta}         {\ensuremath{\Delta\eta}}
\newcommand{\Ntrig}        {\ensuremath{N_{\mathrm{trig}}}}
\newcommand{\Nassoc}       {\ensuremath{N_{\mathrm{assoc}}}}
\newcommand{\dNassoc}      {\ensuremath{\frac{\dd^2N_{\mathrm{assoc}}}{\dd\Deta\dd\Dphi}}}

\newcommand{\Fig}[1]       {Fig.~\ref{#1}}

\newcommand{\Sect}[1]      {Sect.~\ref{#1}}

\newcommand{\Sections}[1]  {Sections~\ref{#1}}
\newcommand{\Eq}[1]        {Eq.~\ref{#1}}

\newcommand{\Ref}[1]       {Ref.~\cite{#1}}

\newcommand{\red}[1]       {\textcolor{red}{#1}}

\newcommand{\warn}[1]      {{\small\textbf{\red{(!}\footnote{\textbf{\red{(!)}}~#1}\red{)}}}\marginpar{\textbf{\red{---}}}}

\newcommand{\final}[1]     {{\textcolor{blue}{#1}}}

\newcommand{\com}[1]       {}

\iflatexdiff

\renewcommand{\xout}[1]    {\textcolor{red}{\sout{#1}}}
 
\else

\renewcommand{\xout}[1]    {}
\fi

\graphicspath{{./img/}}
\ifdraft
\usepackage{lineno}
\linenumbers
\modulolinenumbers[5]
\usepackage{fancyhdr}
\renewcommand{\final}[1]{#1}
\else
\renewcommand{\warn}[1]{}
\renewcommand{\final}[1]{#1}
\fi
\begin{document}
\newlength{\figlen}
\setlength{\figlen}{\linewidth}
\ifpreprint
\setlength{\figlen}{0.95\linewidth}
\begin{titlepage}
\PHnumber{2012-359}                   
\PHdate{03 Dec 2012}                  
\title{\dtitle}
\ShortTitle{\stitle}
\Collaboration{ALICE Collaboration%
         \thanks{See Appendix~\ref{app:collab} for the list of collaboration members}}
\ShortAuthor{ALICE Collaboration} 
\ifdraft
\begin{center}
\today\\ \color{red}DRAFT \dvers\ \hspace{0.3cm} \$Revision: 709 $\color{white}:$\$\color{black}\vspace{0.3cm}
\end{center}
\fi
\else
\begin{frontmatter}
\title{\dtitle}
\iffull
\input{authors-plb.tex}
\else
\ifdraft
\author{ALICE Collaboration \\ \vspace{0.3cm} 
\today\\ \color{red}DRAFT \dvers\ \hspace{0.3cm} \$Revision: 709 $\color{white}:$\$\color{black}}
\else
\author{ALICE Collaboration}
\fi
\fi
\fi
\begin{abstract}
Angular correlations between charged trigger and associated particles are measured 
by the ALICE detector in \pPb\ collisions at a nucleon--nucleon centre-of-mass energy of \unit[5.02]{TeV} 
for transverse momentum ranges within $0.5 < \pta < \ptt < 4$ GeV/$c$.
The correlations are measured over two units of pseudorapidity and full azimuthal angle in different intervals 
of event multiplicity, and expressed as associated yield per trigger particle.
Two long-range ridge-like structures, one on the near side and one 
on the away side, are observed when the per-trigger yield obtained in low-multiplicity events 
is subtracted from the one in high-multiplicity events.
The excess on the near-side is qualitatively similar to that recently reported by the CMS 
collaboration, while the excess on the away-side is reported for the first time. 
The two-ridge structure projected onto azimuthal angle is quantified with the second and third 
Fourier coefficients as well as by near-side and away-side yields and widths. 
The yields on the near side and on the away side are equal within
the uncertainties for all studied event multiplicity and $\pt$ bins, 
and the widths show no significant evolution with event multiplicity or $\pt$. 
These findings suggest that the near-side ridge is accompanied by an essentially 
identical away-side ridge.

\ifdraft 
\ifpreprint
\end{abstract}
\end{titlepage}
\else
\end{abstract}
\end{frontmatter}
\newpage
\fi
\fi
\ifdraft
\thispagestyle{fancyplain}
\else
\end{abstract}
\ifpreprint
\end{titlepage}
\else
\end{frontmatter}
\fi
\fi
\setcounter{page}{2}


\section{Introduction}
\label{sec:intro}
Two-particle correlations are a powerful tool to explore the mechanism of particle 
production in collisions of hadrons and nuclei at high energy.
Such studies involve measuring the distributions of relative angles $\Dphi$ and $\Deta$ between pairs of particles:
 a ``trigger'' particle in a certain transverse momentum $\ptt$ interval and an
``associated'' particle in a $\pta$ interval, where $\Dphi$ and $\Deta$ are the differences in azimuthal angle
$\varphi$ and pseudorapidity $\eta$ between the two particles.

In proton--proton~(\pp) collisions, the correlation at ($\Dphi \approx 0$, $\Deta \approx 0$) for $\ptt>$~\unit[2]{\gevc}
is dominated by the ``near-side'' jet peak, where trigger and associated particles originate from a
fragmenting parton, and at $\Dphi \approx \pi$ by the recoil or ``away-side'' jet~\cite{Wang:1992db}.
The away-side structure is elongated along $\Deta$ due to the longitudinal momentum distribution of
partons in the colliding protons.
In nucleus--nucleus collisions, the jet-related correlations are modified and additional structures
emerge, which persist over a long range in $\Deta$ on the near side and on the away side~\cite{Adams:2004pa,
Alver:2008gk, Alver:2009id, Abelev:2009qa, Chatrchyan:2011eka, Aamodt:2011by, Agakishiev:2011pe,
Chatrchyan:2012wg, ATLAS:2012at, Aamodt:2011vk, Adare:2011tg, Abelev:2012di, Chatrchyan:2012zb}.
The shape of these distributions when decomposed into a Fourier series defined by $v_{n}$
coefficients~\cite{Voloshin:1994mz} is found to be dominated by contributions from terms
with $n=2$ and $n=3$~\cite{Aamodt:2011vk, Adare:2011tg, Aamodt:2011by, Chatrchyan:2011eka, Chatrchyan:2012wg,
Abelev:2012di, Chatrchyan:2012zb, ATLAS:2012at}.
The $v_n$ coefficients are sensitive to the geometry of the initial 
state of the colliding nuclei~\cite{Ollitrault:1992bk,Alver:2010gr} and can be related to the transport properties 
of the strongly-interacting de-confined matter via hydrodynamic models~\cite{Alver:2010dn, Schenke:2010rr, Qiu:2011hf}.

Recently, measurements in \pp\ collisions at a centre-of-mass energy $\s=$~\unit[7]{TeV}~\cite{Khachatryan:2010gv} and 
in proton--lead~(\pPb) collisions at a nucleon--nucleon centre-of-mass energy $\snn=$~\unit[5.02]{TeV} \cite{CMS:2012qk} 
have revealed long-range ($2<|\Deta|<4$) near-side ($\Dphi \approx 0$) correlations in events with significantly 
higher-than-average particle multiplicity.
Various mechanisms have been proposed to explain the origin of these ridge-like correlations in high-multiplicity 
\pp\ and \pPb\ events.
These mechanisms include colour connections forming along the longitudinal direction~\cite{Arbuzov:2011yr,
Dusling:2012cg, Dusling:2012wy, Kovchegov:2012nd}, jet-medium~\cite{Wong:2011qr} and 
multi-parton induced~\cite{Strikman:2011cx, Alderweireldt:2012kt} 
interactions, and collective effects arising in the high-density system possibly formed in these
collisions~\cite{Avsar:2010rf, Werner:2010ss, Deng:2011at, Avsar:2011fz, Bozek:2011if, Bozek:2012gr}.

Results from two-particle correlations in $\snn=$ \unit[0.2]{TeV} \dAu\ 
collisions~\cite{Braidot:2010zh, Adare:2011sc}
show a strong suppression of the away-side yield at forward rapidity in central collisions. 
This modification has been interpreted in the framework of ``Colour Glass Condensate'' 
models~\cite{Albacete:2010pg} as a saturation effect caused by nonlinear gluon interactions 
in the high-density regime at small longitudinal parton momentum fraction $x$. 
Similar effects may arise at midrapidity in \pPb\ collisions at $\snn=$~\unit[5.02]{TeV}, where
the parton distributions are probed down to $x < 10^{-3}$, which is comparable to the
relevant range of $x$ at forward rapidity ($y\sim 3$) at $\snn=$~\unit[0.2]{TeV}.

This letter presents results extracted from two-particle correlation measurements in \pPb\ collisions at $\snn=$~\unit[5.02]{TeV}, 
recorded with the \mbox{ALICE} 
detector~\cite{Aamodt:2008zz} at the Large Hadron Collider~(LHC).
The correlations are measured over two units of pseudorapidity and full azimuthal angle as a function 
of charged-particle multiplicity, and expressed as associated yield per trigger particle.
\Sections{sec:setup} and \ref{sec:selection} describe the experimental setup,
and the event and track selection, respectively.
Details on the definition of the correlation and the per-trigger-particle associated yield
are given in \Sect{sec:twopartfunc}.
The results of the analysis are discussed in \Sect{sec:results} and a 
summary is given in \Sect{sec:summary}.

\section{Experimental setup}
\label{sec:setup}
Collisions of proton and lead beams were provided by the LHC during a short 
pilot run performed in September 2012.
The beam energies were \unit[4]{TeV} for the proton beam and \unit[1.58]{TeV} per nucleon for the lead beam,
resulting in collisions at $\snn=$~\unit[5.02]{TeV}. The nucleon--nucleon centre-of-mass 
system moved with respect to the ALICE laboratory system with a rapidity of $-0.465$, i.e.,
in the direction of the proton beam.
The pseudorapidity in the laboratory system is denoted with $\eta$ throughout this letter.
Results from \pp\ collisions at $\s=$ 2.76 and \unit[7]{TeV}
are shown in comparison to the \pPb\ results.

A detailed description of the ALICE detector can be found in \Ref{Aamodt:2008zz}.
The main subsystems used in the present analysis are the Inner Tracking System~(\ITS) and the
Time Projection Chamber~(\TPC), which are operated 
inside a solenoidal magnetic field of \unit[0.5]{T}. 
The \ITS\ consists of six layers of silicon detectors: from the innermost to the outermost, 
two layers of Silicon Pixel Detector (\SPD) with an acceptance of $|\eta| < 1.4$, 
two layers of Silicon Drift Detector (\SDD) with $|\eta| < 0.9$ and two layers of Silicon Strip 
Detector with $|\eta| < 0.97$.
The \TPC\ provides an acceptance of $|\eta|<0.9$ for tracks which reach the outer radius of the \TPC\ and up to $|\eta|<1.5$
for tracks with reduced track length.
The VZERO detector, two arrays of 32 scintillator tiles each, covering the full azimuth within
$2.8<\eta<5.1$~(\VZEROA) and $-3.7<\eta<-1.7$~(\VZEROC), was used for triggering, 
event selection and 
event characterization, namely the definition of event classes corresponding to different
particle-multiplicity ranges. In \pPb\ collisions, the trigger required a signal in either \VZEROA\ or \VZEROC.
In addition, two neutron Zero Degree Calorimeters~(\ZDCs) located at \final{$+112.5$}~m~(\ZNA) and
\final{$-112.5$}~m~(\ZNC) from the interaction point are used in the event selection.
The energy deposited in the \ZNA, which for the beam setup of the pilot run originates from neutrons of 
the Pb nucleus,
served as an alternative approach in defining the event-multiplicity classes.
In pp collisions, the trigger required a signal in either \SPD, \VZEROA\ or \VZEROC\ \cite{Aamodt:2010ft}.

\section{Event and track selection}
\label{sec:selection}
The present analysis of the \pPb\ data is based on the event selection described in \Ref{ALICE:2012xs}.
The events are selected by requiring a signal in both \VZEROA\ and \VZEROC.
From the data collected, $1.7\times 10^6$ events pass the event selection criteria and are used for this analysis. 
For the analysis of the \pp\ collisions, the event selection described in \Ref{Aamodt:2010ft} has been used, 
yielding $31 \times 10^6$ and $85 \times 10^6$ events at $\s=$ 2.76 and \unit[7]{TeV}, 
respectively.

The primary-vertex position is determined with tracks reconstructed in the \ITS\ and \TPC\ as described 
in \Ref{Abelev:2012eq}. The vertex reconstruction algorithm is fully efficient for events with at least 
one reconstructed primary track within $|\eta|<1.4$~\cite{perfpaper}.
An event is accepted if the coordinate of the reconstructed vertex along the beam
direction ($z_{\rm vtx}$) is within \mbox{$\pm10$~cm} from the detector centre.

The analysis uses tracks reconstructed in the \ITS\ and \TPC\ with $0.5<\pt<$~\unit[4]{\gevc} and in a fiducial region $|\eta|<1.2$. 
As a first step in the track selection, cuts on the number of space points and the quality of the track fit 
in the \TPC\ are applied. 
Tracks are further required to have a distance of closest approach to the reconstructed vertex smaller than 
\unit[2.4]{cm} and \unit[3.2]{cm} in the transverse and the longitudinal direction, respectively.
In order to avoid an azimuthally-dependent tracking efficiency due to inactive \SPD\ modules, two classes of tracks 
are combined~\cite{Abelev:2012ej}. 
The first class consists of tracks, which have at least one hit in the \SPD. 
The tracks from the second class do not have any \SPD\ associated hit, but the position of the reconstructed 
primary vertex is used in the fit of the tracks.
In the study of systematic uncertainties an alternative track selection \cite{ALICE:2011ac} is used, where
a tighter $\pt$-dependent cut on the distance of closest approach to the reconstructed vertex is applied. 
Further, the selection for the tracks in the second class is changed to tracks, which have a 
hit in the first layer of the \SDD. 
This modified selection has a less uniform azimuthal acceptance, but includes a smaller number of secondary particles 
from interactions in the detector material or weak decays.

The efficiency and purity of the primary charged-particle selection are estimated from a Monte Carlo~(MC)
simulation using the DPMJET event generator~\cite{Roesler:2000he} (for \pPb) and the PYTHIA~6.4 event 
generator~\cite{Sjostrand:2006za} with the tune Perugia-0~\cite{Skands:2010ak} (for \pp) with particle transport 
through the detector using GEANT3~\cite{geant3ref2}.
In \pPb\ collisions, the combined efficiency and acceptance for the track reconstruction in $|\eta|<0.9$ is 
about 82\% at $\pt=$~\unit[0.5--1]{\gevc}, and decreases to about $79$\% at $\pt=$~\unit[4]{\gevc}. 
It reduces to about $50$\% at $|\eta| \approx 1.2$ and 
is independent of the event multiplicity.
The remaining contamination from secondary particles due to interactions in the detector material 
or weak decays decreases from about 2\% to 1\% in the $\pt$ range from 0.5 to \unit[4]{\gevc}. 
The contribution from fake tracks is negligible. 
These fractions are similar in the analysis of \pp\ collisions.

In order to study the multiplicity dependence of the two-particle correlations the selected event sample is 
divided into four event classes. 
These classes are defined fractions of the analyzed event sample, based on cuts on the total charge 
deposited in the VZERO detector (V0M), and denoted ``60--100\%", ``40--60\%", ``20--40\%", ``0--20\%" from the lowest to the 
highest multiplicity in the following.
Table~\ref{tab:multclasses} shows the event-class definitions and the 
corresponding mean charged-particle multiplicity densities~($\avg{\dNdeta}$) within $|\eta|<0.5$.
These are obtained using the method presented in Ref.~\cite{ALICE:2012xs},
and are corrected for acceptance and tracking efficiency as well as contamination by secondary particles.
Also shown are the mean numbers of primary charged particles with $\pt>$~\unit[0.5]{\gevc} within the 
pseudorapidity range $|\eta|<1.2$.
These are measured by applying the track selection described above and are
corrected for the detector acceptance, track-reconstruction efficiency and contamination. 

\begin{table}[bht!f] \centering
  \begin{tabular}{cccc}
    \hline
    Event        & V0M range & $\avg{\dNdeta}|_{|\eta|<0.5}$  & $\avg{N_{\rm trk}}|_{|\eta|<1.2}$ \\
    class        & (a.u.)    & $\pt>$~\unit[0]{\gevc} & $\pt>$~\unit[0.5]{\gevc} \\
    \hline
      60--100\% & $<138$    & $ 6.6 \pm 0.2$                & $6.4 \pm 0.2$ \\
      40--60\%  & 138--216  & $16.2 \pm 0.4$               & $16.9 \pm 0.6$ \\
      20--40\%  & 216--318  & $23.7 \pm 0.5$               & $26.1 \pm 0.9$ \\
       0--20\%  & $>318$    & $34.9 \pm 0.5$               & $42.5 \pm 1.5$ \\
    \hline
  \end{tabular}
  \caption{\label{tab:multclasses}
    Definition of the event classes as fractions of the analyzed event sample and their corresponding $\avg{\dNdeta}$ within $|\eta|<0.5$
    and the mean numbers of charged particles within $|\eta|<1.2$ and $\pt>$~\unit[0.5]{\gevc}. 
    The given uncertainties are systematic as the statistical uncertainties are negligible.}
\end{table}

\section{Analysis}
\label{sec:twopartfunc}
For a given event class,
the two-particle correlation between pairs of trigger and associated charged
particles is measured as a function of the azimuthal difference $\Dphi$ (defined within $-\pi/2$ and $3\pi/2$) and pseudorapidity
difference $\Deta$. The correlation is expressed 
in terms of the associated yield per trigger particle for different intervals of trigger and associated transverse momentum, 
$\ptt$ and $\pta$, respectively, and $\pta < \ptt$.
The associated yield per trigger particle is defined as
\begin{equation}
\frac{1}{\Ntrig} \dNassoc = \frac{S(\Deta,\Dphi)}{B(\Deta,\Dphi)} \label{pertriggeryield}
\end{equation}
where $\Ntrig$ is the total number of trigger particles in the event class and $\ptt$ interval.
The signal distribution 
$S(\Deta,\Dphi) = 1/\Ntrig \dd^2N_{\rm same}/\dd\Deta\dd\Dphi$
is the associated yield per trigger particle for particle pairs from the same event. 
In a given event class and $\pt$ interval, the sum over the events is performed separately for $\Ntrig$ and 
$\dd^2N_{\rm same}/\dd\Deta\dd\Dphi$ before their ratio is computed. 
Note, that this definition is different from the one used in \Ref{CMS:2012qk}, where $S(\Deta,\Dphi)$ is calculated 
per event and then averaged. 
The method used in this letter does not induce an inherent multiplicity dependence in the pair yields, which is important
for the subtraction method discussed in the next Section.
The background distribution $B(\Deta,\Dphi) = \alpha\ \dd^2N_{\rm mixed}/\dd\Deta\dd\Dphi$
corrects for pair acceptance and pair efficiency.
It is constructed by correlating the trigger particles in one event with the associated particles from
other events in the same event class and within the same \unit[2]{cm} wide $z_{\rm vtx}$ interval 
(each event is mixed with 5--20 events).
The factor $\alpha$ is chosen to normalize the background distribution such that it is unity for pairs where
both particles go into approximately the same direction (i.e.\ $\Dphi\approx 0,\Deta\approx 0$).
To account for different pair acceptance and pair efficiency as a function of $z_{\rm vtx}$, the yield defined by
\Eq{pertriggeryield} is constructed for each $z_{\rm vtx}$ interval.
The final per-trigger yield is obtained by calculating the weighted average of the $z_{\rm vtx}$ intervals.

When constructing the signal and background distributions, the trigger and associated particles are required to be separated 
by $|\Delta\varphi^{*}_{\rm min}|>0.02$ and $|\Deta|>0.02$, 
where $\Delta\varphi^*_{\rm min}$ is the minimal azimuthal distance at the same radius between the two tracks within
the active detector volume after accounting for the bending due to the magnetic field. 
This procedure is applied to avoid a bias due to the reduced efficiency for pairs with small opening angles 
and leads to an increase in the associated near-side peak yield of 0.4--0.8\% depending on $\pt$.
Furthermore, particle pairs are removed which are likely to stem from a $\gamma$-conversion, or a $K^0_s$ or 
$\Lambda$ decay, by a cut on the invariant mass of the pair (the electron, pion, or pion/proton mass is assumed, respectively). 
The effect on the near-side peak yields is less than 2\%.

In the signal as well as in the background distribution, each trigger and each associated particle is weighted with a correction 
factor that accounts for detector acceptance, reconstruction efficiency and contamination by secondary particles.
These corrections are applied as a function of $\eta$, $\pt$ and $z_{\rm vtx}$.
Applying the correction factors extracted from DPMJET simulations to events simulated with HIJING~\cite{hijing} leads 
to associated peak yields that agree within 4\% with the MC truth. 
This difference between the two-dimensional corrected per-trigger yield and input per-trigger yield is used in the estimate 
of the systematic uncertainties.
Uncertainties due to track-quality cuts are evaluated by comparing the results of two different track selections, 
see \Sect{sec:selection}. 
The associated yields are found to be insensitive to these track selections within 5\%.
Further systematic uncertainties related to specific observables are mentioned below.

\begin{figure}[hbt!f]
\centering
\includegraphics[width=0.49\textwidth]{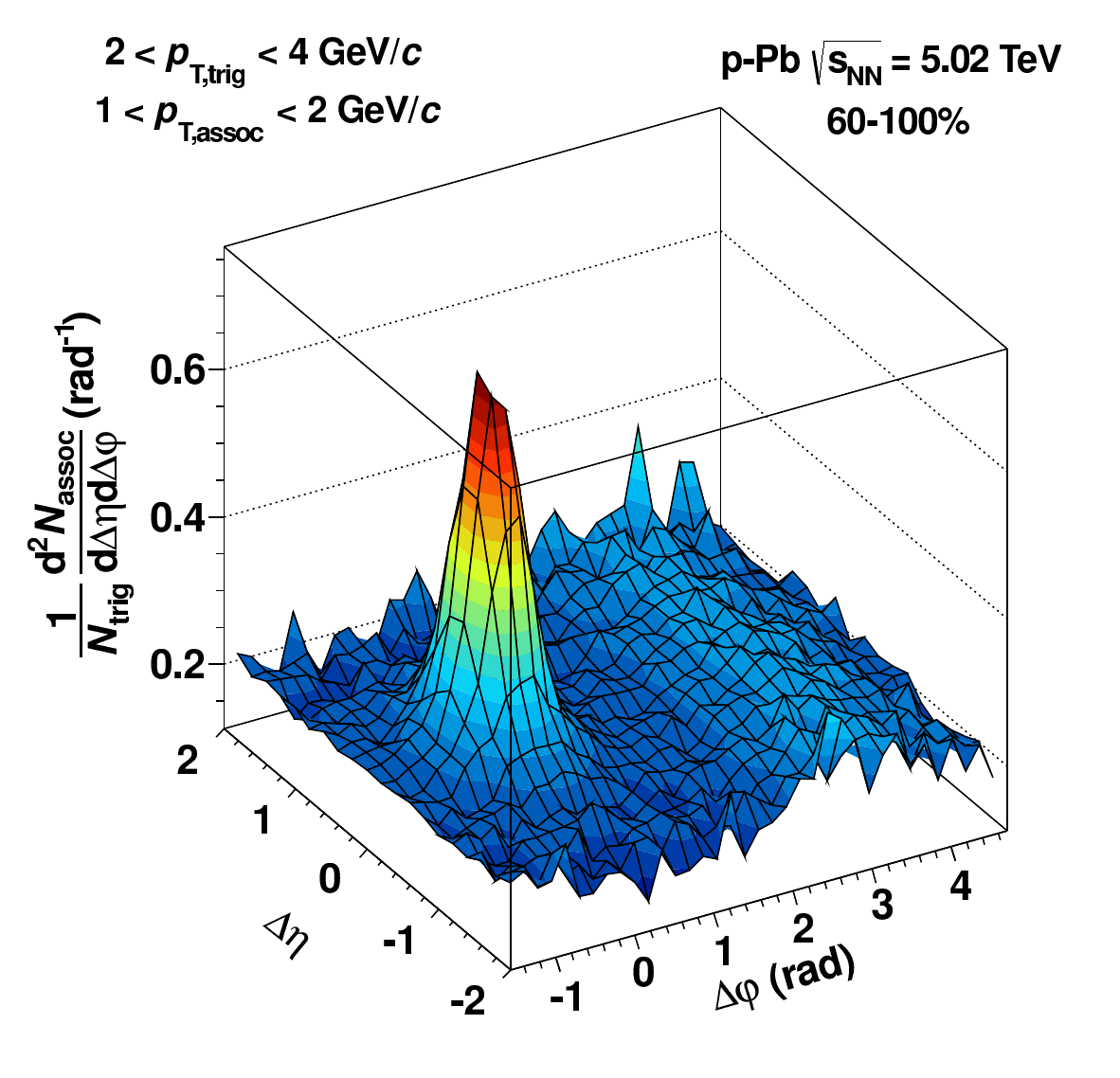}
\hfill
\includegraphics[width=0.49\textwidth]{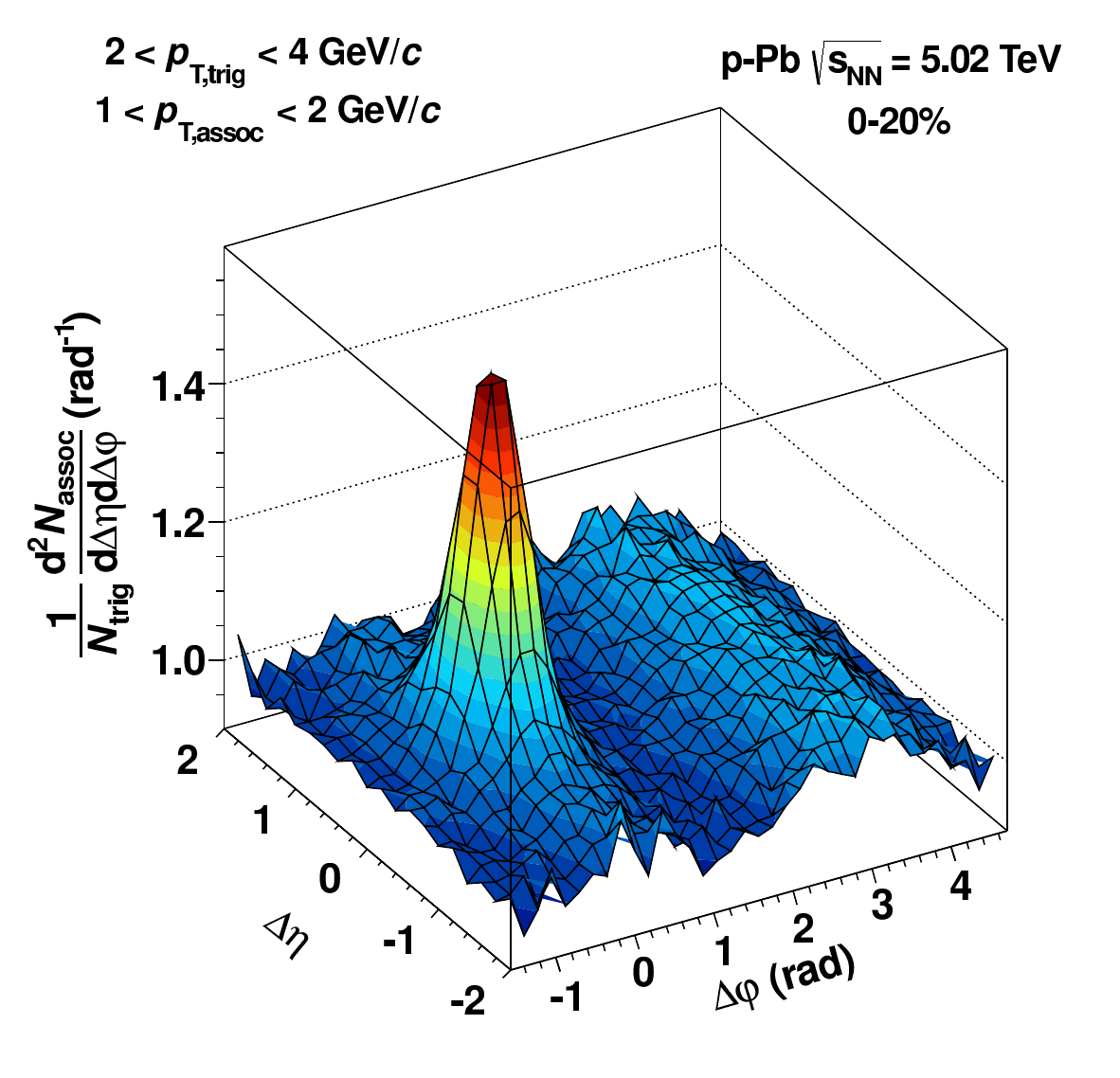}
\caption{\label{fig:2dcorr}
The associated yield per trigger particle in $\Dphi$ and $\Deta$ 
for pairs of charged particles with $2<\ptt<$~\unit[4]{\gevc} and $1<\pta<$~\unit[2]{\gevc} in \pPb\ collisions 
at $\snn=5.02$ TeV for the 60--100\% (left) and 0--20\% (right) event classes.
}
\end{figure}

\section{Results}
\label{sec:results}
\begin{figure}[thbt!f]
\centering
\includegraphics[width=0.7\textwidth]{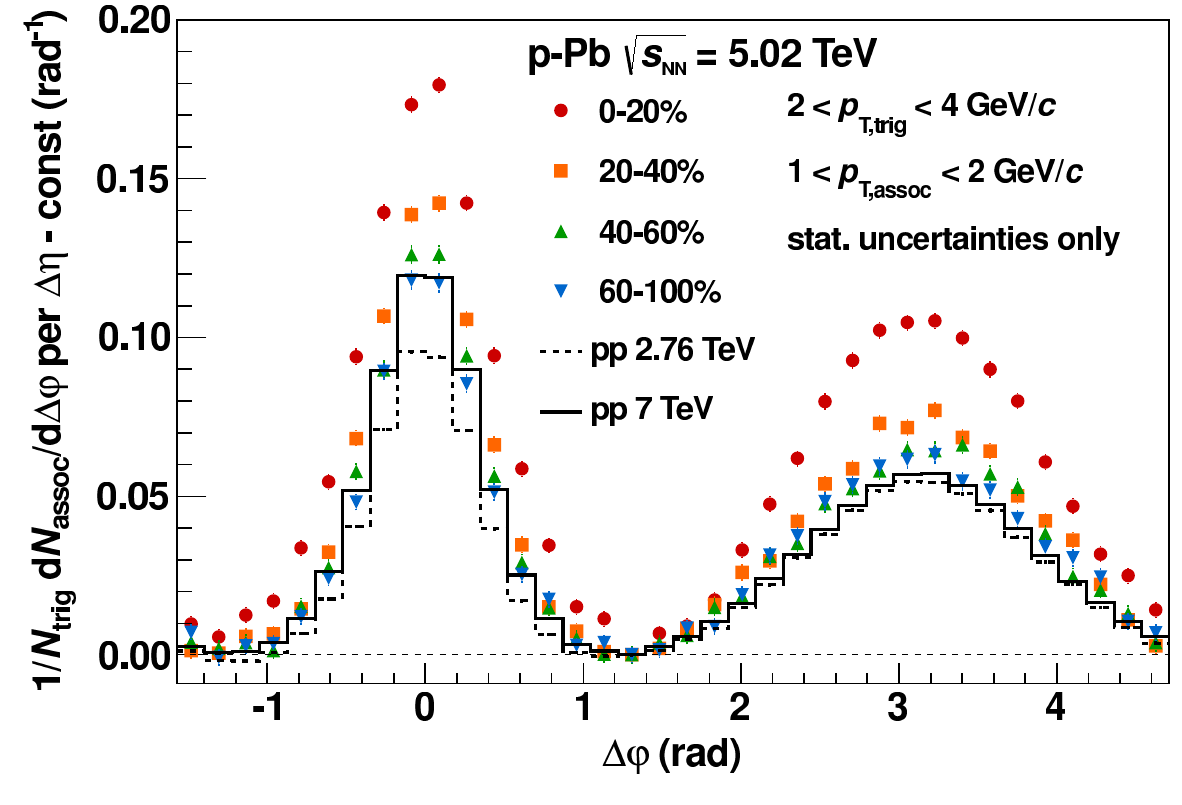}
\caption{\label{fig:1dcorr}
Associated yield per trigger particle
as a function of $\Dphi$ averaged over $|\Deta| < 1.8$ for pairs of charged particles
with $2<\ptt<$~\unit[4]{\gevc} and $1<\pta<$~\unit[2]{\gevc} in \pPb\ collisions at $\snn=5.02$ TeV for
different event classes, and in \pp\ collisions at 2.76 and 7 TeV. 
The yield between the peaks (determined at $\Dphi \approx 1.3$) has been subtracted in each case. Only statistical uncertainties 
are shown; systematic uncertainties are less than 0.01 (absolute) per bin.
}
\end{figure}

The associated yield per 
trigger particle in $\Dphi$ and $\Deta$ is shown in \Fig{fig:2dcorr} for pairs of charged 
particles with $2<\ptt<$~\unit[4]{\gevc} and $1<\pta<$~\unit[2]{\gevc} in \pPb\ collisions at $\snn=$ \unit[5.02]{TeV} 
in the 60--100\% (left) and 0--20\% (right) event classes.
In the 60--100\% class, the visible features are the correlation peak near
$(\Dphi\approx 0,\Deta\approx 0)$ for pairs of particles originating from the same jet, and the
elongated structure at $\Dphi \approx \pi$ for pairs of particles back-to-back in azimuth. 
These are similar to those observed in \pp\ collisions at $\s = 2.76$ and \unit[7]{TeV}.
The same features are visible in the 0--20\% class. However, both the yields on the near side ($|\Dphi| < \pi/2$)
and the away side ($\pi/2 < \Dphi < 3\pi/2$) are higher.~\footnote{These definitions of near-side ($|\Dphi| < \pi/2$)
and away-side ($\pi/2 < \Dphi < 3\pi/2$) are used throughout the letter.} 
This is illustrated in \Fig{fig:1dcorr}, where the projections
on $\Dphi$ averaged over $|\Deta| < 1.8$ are compared for different event classes and also compared to \pp\ 
collisions at 2.76 and \unit[7]{TeV}.
In order to facilitate the comparison, the yield at $\Dphi=1.3$ has been subtracted for each distribution.
It is seen that the per-trigger yields in $\Dphi$ on the near side and on the away side are similar for
low-multiplicity \pPb\ collisions and for \pp\ collisions at $\s=$~\unit[7]{TeV}, and increase with increasing 
multiplicity in \pPb\ collisions.

\begin{figure}[thb!f]
\centering
 \begin{minipage}[b]{0.54\textwidth}
 \includegraphics[width=\linewidth]{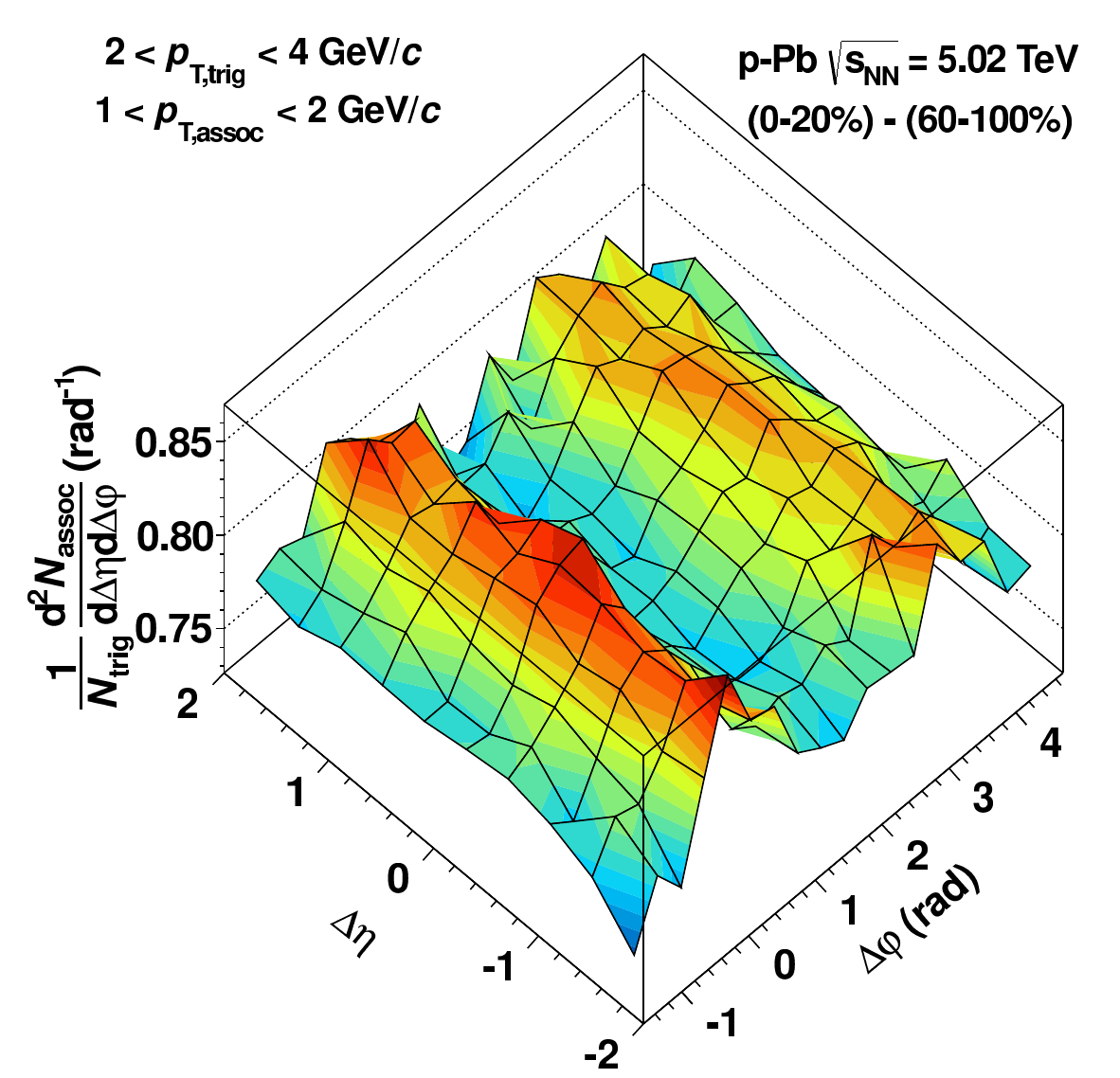}
 \end{minipage}
 \begin{minipage}[b]{0.45\textwidth}
  \begin{minipage}[t]{\textwidth}
  \includegraphics[width=0.9\linewidth]{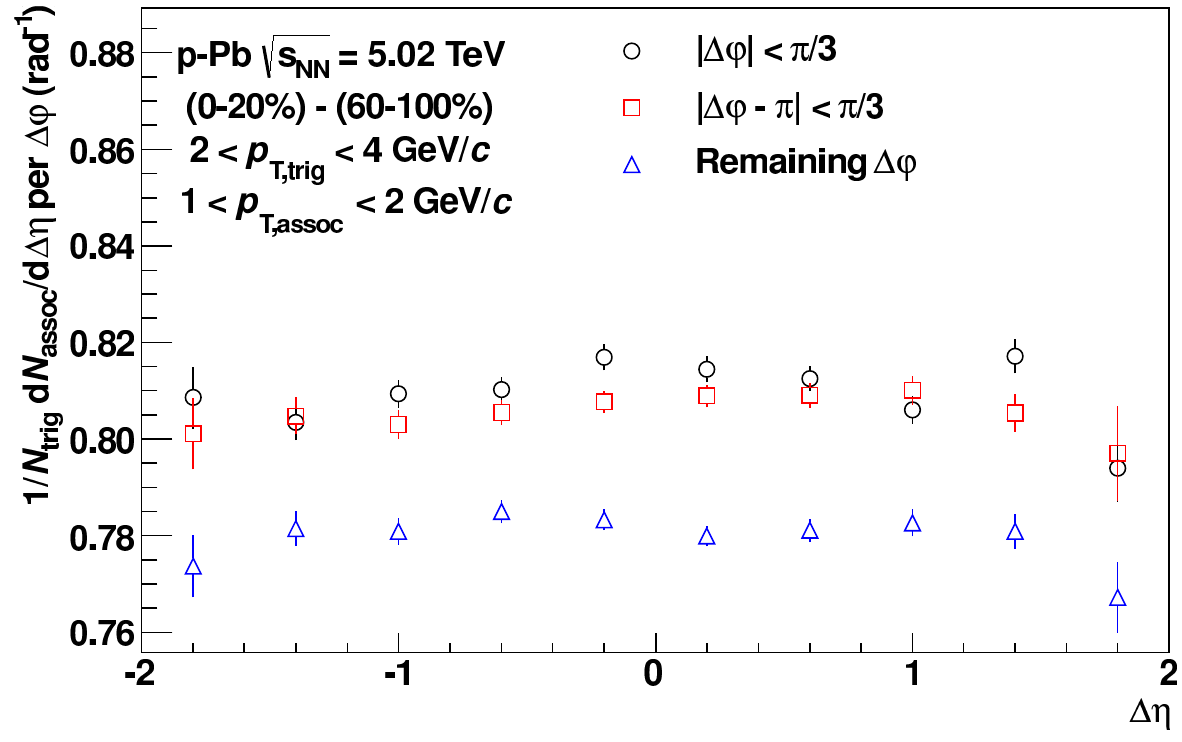}
  \end{minipage}
 \hfill
  \begin{minipage}[t]{\textwidth}
  \includegraphics[width=0.9\linewidth]{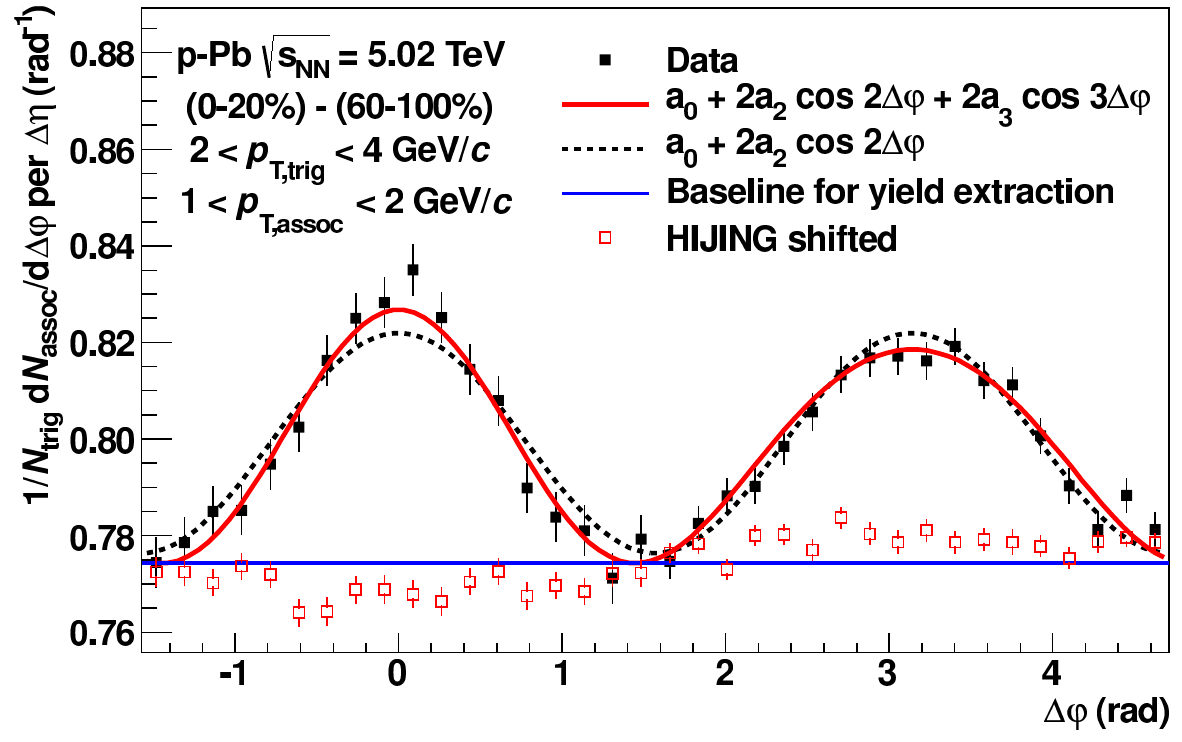}
  \end{minipage}
 \end{minipage}
\caption{\label{fig:centminper}
Left: Associated yield per trigger particle in $\Dphi$ and $\Deta$ for pairs of 
charged particles with $2<\ptt<4$ GeV/$c$ and $1<\pta<2$ GeV/$c$ in \pPb\ collisions at $\snn=5.02$ TeV for the 0--20\%
multiplicity class, after subtraction of the associated yield obtained in the 60--100\% event class.
Top right: the associated per-trigger yield after subtraction (as shown on the left) projected onto 
$\Deta$ averaged over $|\Dphi| < \pi/3$ (black circles), $|\Dphi - \pi| < \pi/3$ (red squares), 
and the remaining area (blue triangles, $\Dphi < -\pi/3$, $\pi/3 < \Dphi < 2\pi/3$ and $\Dphi > 4\pi/3$).
Bottom right: as above but projected onto $\Dphi$ averaged over $0.8 < |\Deta| < 1.8$ 
on the near side and $|\Deta| < 1.8$ on the away side.
Superimposed are fits containing a $\cos(2\Dphi)$ shape alone (black dashed line) and a combination of $\cos(2\Dphi)$
and $\cos(3\Dphi)$ shapes (red solid line).
The blue horizontal line shows the baseline obtained from the latter fit which is used for the yield calculation. 
Also shown for comparison is the subtracted associated yield when the same procedure is applied on HIJING shifted 
to the same baseline.
The figure shows only statistical uncertainties. 
Systematic uncertainties are mostly correlated and affect the baseline. 
Uncorrelated uncertainties are less than 1\%.
}
\end{figure}

To quantify the change from low to high multiplicity event classes, we subtract the per-trigger yield
of the lowest (60--100\%) from that of the higher multiplicity classes.
The resulting distribution in $\Dphi$ and $\Deta$ for the 0--20\% 
event class is shown in \Fig{fig:centminper}~(left).
A distinct excess structure in the correlation is observed,
which forms two ridges, one on the near side and one on the away side.
The ridge on the near side is qualitatively similar to the one recently reported by the CMS
collaboration~\cite{CMS:2012qk}.  
Note, however that a quantitative comparison would not be meaningful due to the different definition of the per-trigger yield 
and the different detector acceptance and event-class definition.

On the near side, there is a peak around ($\Dphi \approx 0$, $\Deta \approx 0$) indicating a small change of the near-side 
jet yield as a function of multiplicity. 
The integral of this peak above the ridge within $|\Deta| < 0.5$ corresponds to about 5--25\% of the unsubtracted near-side peak yield, 
depending on $\pt$.
In order to avoid a bias on the associated yields due to the multiplicity selection and to prevent
that this remaining peak contributes to the ridge yields calculated below, the region $|\Deta| < 0.8$
on the near side is excluded when performing projections onto $\Dphi$.
The effect of this incomplete subtraction on the extracted observables, which if jet-related might also be present on the 
away side, is discussed further below.

The top right panel in~\Fig{fig:centminper} shows the projection of 
\Fig{fig:centminper}~(left) onto $\Deta$ averaged over different $\Dphi$ intervals. 
The near-side and away-side distributions are flat apart from the discussed small peak around $\Deta = 0$. 
The bottom right panel shows the projection to $\Dphi$, where a modulation is observed.
For comparison, the subtracted associated yield for HIJING simulated events shifted to 
the baseline of the data is also shown, where no significant modulation remains.
To quantify the near-side and away-side excess structures, the following functional form
\begin{equation}
  1/\Ntrig \dd \Nassoc/\dd\Dphi = a_0 + 2\,a_2 \cos(2\Dphi) + 2\,a_3 \cos(3\Dphi)
  \label{fitfunction}
\end{equation}
is fit to the data in multiplicity and $\pt$ intervals. 
The fits have a $\chi^2/{\rm ndf}$ of less than $1.5$ with and less than $1.8$ without the $a_3 \cos(3\Dphi)$ term 
in the different $\pt$ and multiplicity intervals, indicating that the data are well described by the fits. 
An example for the fit with and without the $a_3 \cos(3\Dphi)$ term is shown
in the bottom right panel of \Fig{fig:centminper}. 
The fit parameters $a_2$ and $a_3$ are a measure of the absolute modulation in the subtracted per-trigger yield 
and characterize a modulation relative to the baseline $b$ in the higher multiplicity class assuming that such a modulation 
is not present in the 60--100\% event class. 
This assumption has been checked by subtracting the yields obtained in $\s = 2.76$ and \unit[7]{TeV} \pp\ collisions 
from the yields obtained for the 60--100\% \pPb\ event class and verifying that in both cases no significant signal remains.
Therefore, the Fourier coefficients $v_n$ of the corresponding single-particle distribution, 
commonly used in the analysis of particle correlations in nucleus--nucleus collisions~\cite{Voloshin:1994mz}, can be 
obtained in bins where the $\ptt$ and $\pta$ intervals are identical using
\begin{equation}
  v_n = \sqrt{a_n / b}.
  \label{vn}
\end{equation}
The baseline $b$ is evaluated in the higher-multiplicity class in the region $|\Dphi-\pi/2|< 0.2$, corrected for the fact 
that it is obtained in the minimum of \Eq{fitfunction}.
A potential bias due to the above-mentioned incomplete near-side peak subtraction on $v_2$ and $v_3$ 
is evaluated in the following way: 
a) the size of the near-side exclusion region is changed from $|\Deta| < 0.8$ to $|\Deta| < 1.2$; b)
the residual near-side peak above the ridge is also subtracted from the away side by mirroring it at $\Dphi = \pi/2$ 
accounting for the general $\pt$-dependent difference of near-side and away-side jet yields due to the kinematic constraints 
and the detector acceptance, which is evaluated using the lowest multiplicity class; and c)
the lower multiplicity class is scaled before the subtraction such that no residual near-side peak above the ridge remains. 
The resulting differences in $v_2$ (up to 15\%) and $v_3$ coefficients (up to 40\%) when applying these approaches 
have been added to the systematic uncertainties. 

\begin{figure}[hbt!f]
\centering
\includegraphics[width=0.49\textwidth]{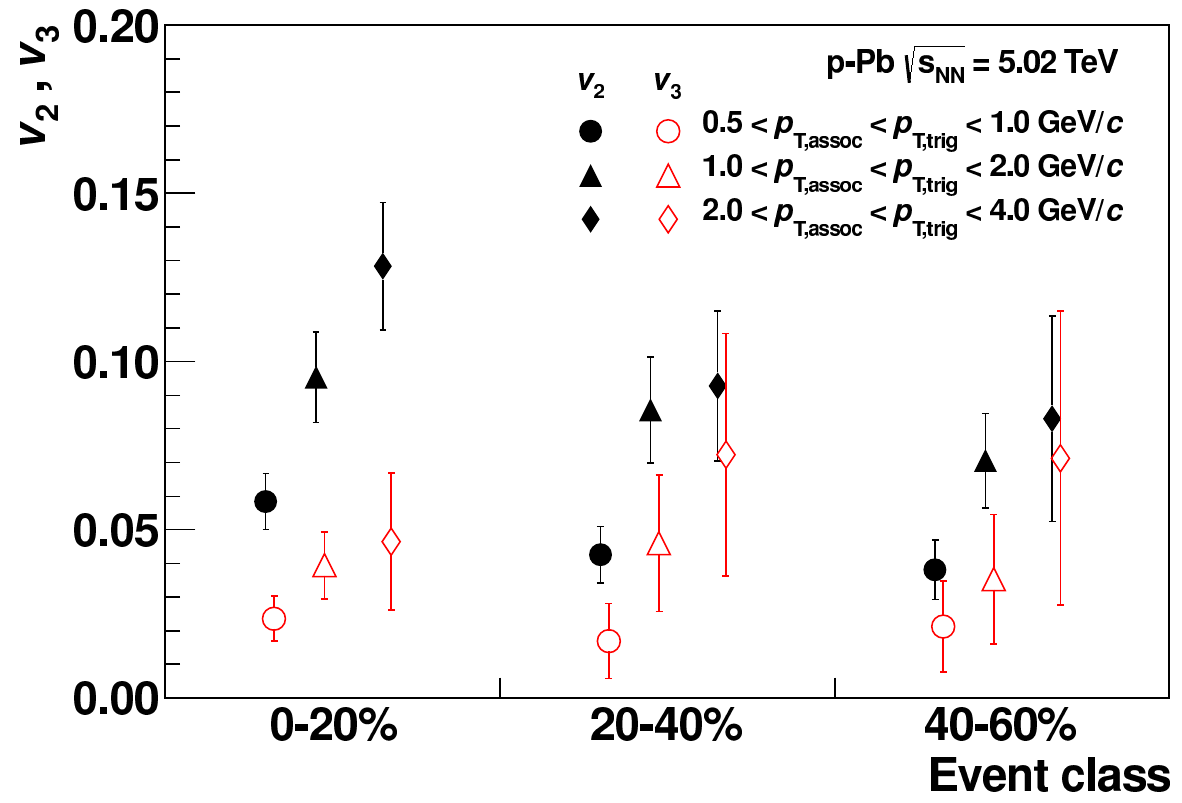}
\hfill
\includegraphics[width=0.49\textwidth]{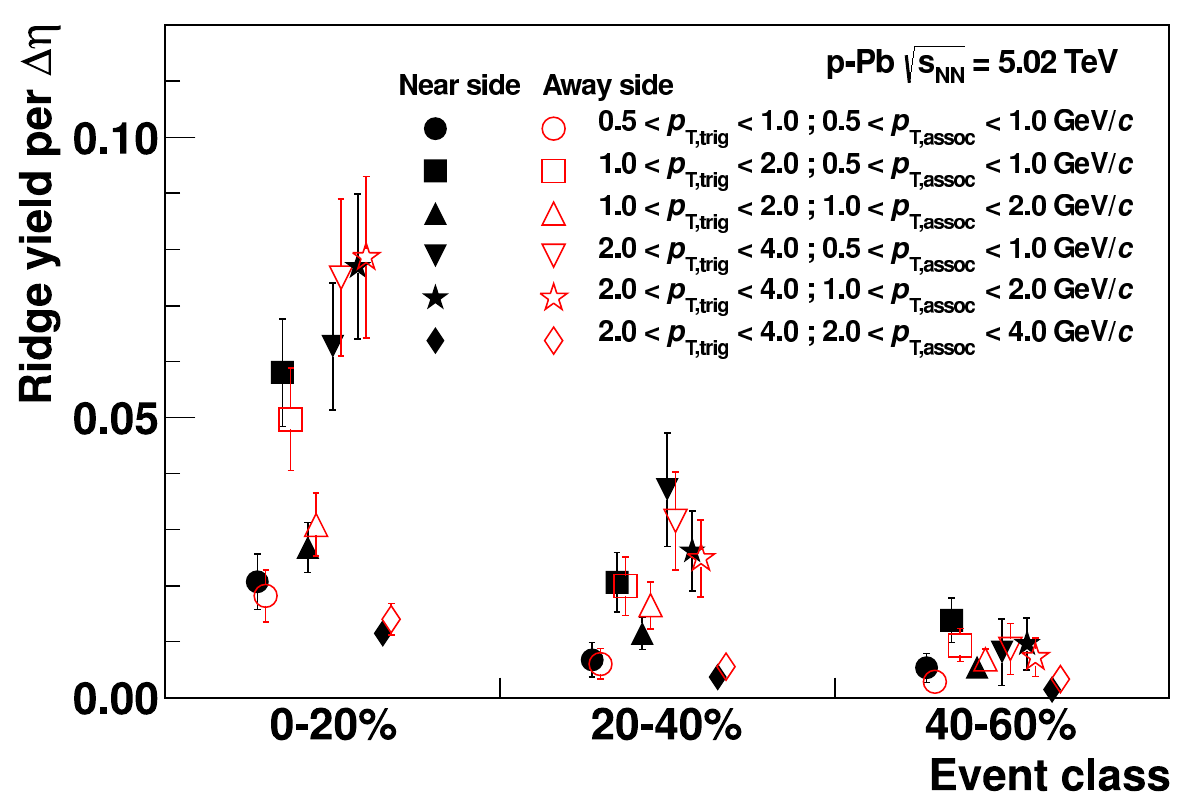}
\caption{\label{fig:v2andyields}
Left: $v_2$ (black closed symbols) and $v_3$ (red open symbols) for different multiplicity
classes and overlapping $\pta$ and $\ptt$ intervals.
Right: Near-side (black closed symbols) and away-side (red open symbols) ridge yields per unit 
of $\Deta$ for different $\ptt$ and $\pta$
bins as a function of the multiplicity class. The error bars show statistical and systematic uncertainties added in quadrature. 
In both panels the points are slightly displaced horizontally for visibility.
}
\end{figure}

The coefficients $v_2$ and $v_3$ are shown in the left panel of \Fig{fig:v2andyields} for different event classes.
The coefficient $v_2$ increases with increasing $\pt$, and shows only a small dependence on multiplicity. 
In the 0--20\% event class, $v_2$  increases from $0.06 \pm 0.01$ for $0.5<\pt<$~\unit[1]{\gevc} to 
$0.12\pm0.02$ for $2<\pt<$~\unit[4]{\gevc}, 
while $v_3$ is about 0.03 and shows, within large errors, an increasing trend with $\pt$.
Reference~\cite{Bozek:2011if} gives predictions for two-particle correlations arising from collective flow in \pPb\ collisions at the 
LHC in the framework of a hydrodynamical model. 
The values for $v_2$ and $v_3$ coefficients, as well as the $\pt$ and the multiplicity dependences, 
are in qualitative agreement with the presented results.

To extract information on the yields and widths of the excess distributions in \Fig{fig:centminper} (bottom right), 
a constant baseline assuming zero yield at the minimum of the fit function (Eq.~\ref{fitfunction}) is subtracted. 
The remaining yield is integrated on the near side and on the away side. Alternatively, a baseline evaluated from the minimum 
of a parabolic function fitted within $|\Dphi - \pi/2| < 1$ is used;
the difference on the extracted yields is added to the systematic uncertainties. 
The uncertainty imposed by the residual near-side jet peak on the yield
 is evaluated in the same way as for the $v_n$ coefficients.
The near-side and away-side ridge yields are shown in the right panel of
\Fig{fig:v2andyields} for different event classes and for different combinations of
$\ptt$ and $\pta$ intervals.
The near-side and away-side yields range from 0 to 0.08 per unit of $\Deta$ depending on multiplicity class and $\pt$ interval. 
It is remarkable that the near-side and away-side yields always agree within uncertainties for a given sample
despite the fact that the absolute values change substantially with event class and $\pt$ interval.
Such a tight correlation between the yields is non-trivial 
and suggests a common underlying physical origin for the near-side and the away-side ridges.

From the baseline-subtracted per-trigger yields the square root of the variance, 
$\sigma$, within $|\Dphi| < \pi/2$ and $\pi/2 < \Dphi < 3\pi/2$ for the near-side and away-side region, respectively, is calculated.
The extracted widths on the near side and the away side agree with each other within 20\% and vary between 0.5 and 0.7. 
There is no significant $\pt$ dependence, which suggests that the observed ridge is not of jet origin.

\begin{figure}[thbt!f]
\centering
\includegraphics[width=0.7\textwidth]{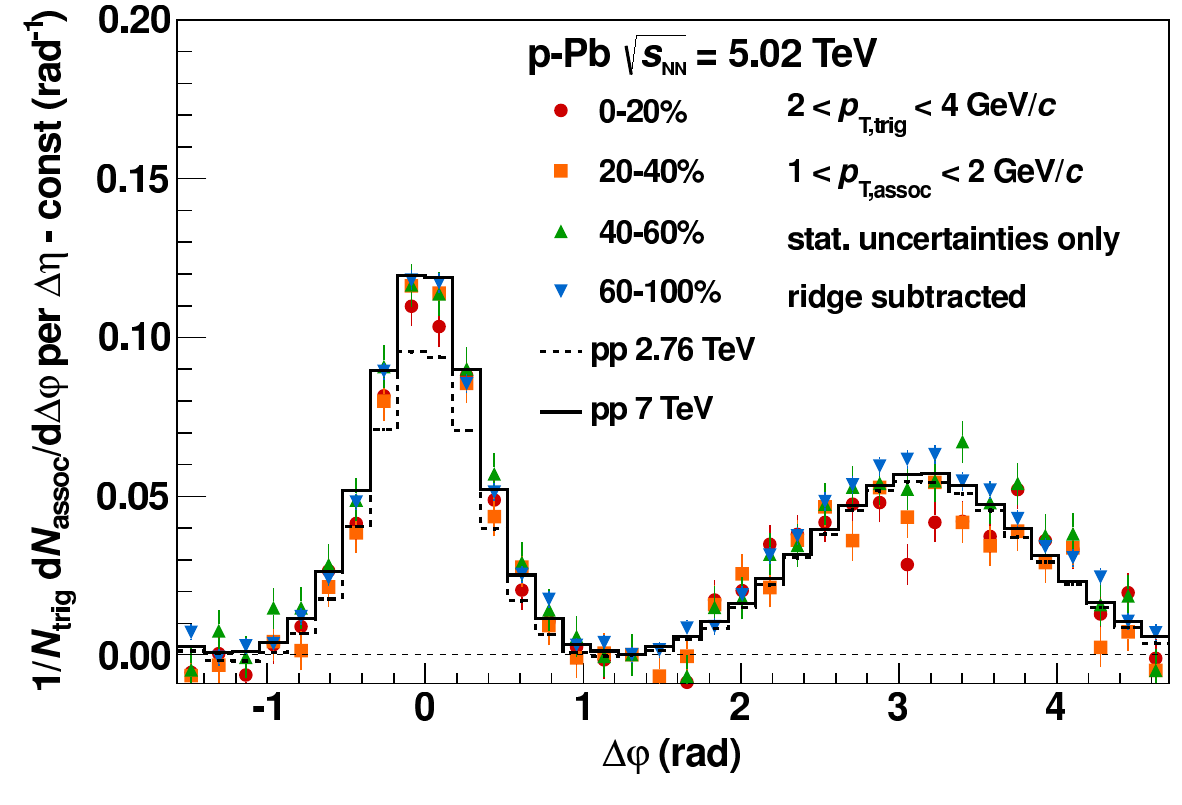}
\caption{\label{fig:1dcorr_ridgesubtracted} 
Associated yield per trigger particle as a function of $\Dphi$ averaged over $|\Deta| < 1.8$ for pairs of charged particles
with $2<\ptt<$~\unit[4]{\gevc} and $1<\pta<$~\unit[2]{\gevc} in \pPb\ collisions at $\snn=5.02$ TeV for
different event classes, compared to \pp\ collisions at $\s = 2.76$ and \unit[7]{TeV}.
For the event classes 0--20\%, 20--40\% and 40--60\% the long-range contribution on the near-side
$1.2<|\Deta|<1.8$ and $|\Dphi|<\pi/2$ has been subtracted from both the near side
and the away side as described in the text. 
Subsequently, the yield between the peaks (determined at $\Dphi \approx 1.3$) has been subtracted 
in each case. 
Only statistical uncertainties are shown; systematic uncertainties are less than 0.01 (absolute) per bin.
}
\end{figure}

The analysis has been repeated using the forward \ZNA\ detector instead of the \VZERO\ for the definition of the event classes.
Unlike in nucleus--nucleus collisions, the correlation between forward energy measured in the \ZNA\ and particle density at central 
rapidities is very weak in proton--nucleus collisions. 
Therefore, event classes defined as fixed fractions of the signal distribution in the \ZNA\ select different events, with different mean 
particle multiplicity at midrapidity, than the samples selected with the same fractions in the \VZERO\ detector.
While the event classes selected with the \ZNA\ span a much smaller range in central multiplicity density, they also minimize any 
autocorrelation between multiplicity selections and, for example, jet activity.
With the \ZNA\ selection, we find qualitatively consistent results compared to the VZERO selection.
In particular, an excess in the difference between low-multiplicity and high-multiplicity \ZNA\ selected events is observed 
to be symmetric on the near side and away side.
Also, the $v_n$ coefficients and $\sigma$ widths are similar within uncertainties. 
However, both the ridge yields and mean charged-particle multiplicity density at midrapidity are different 
between the \VZERO\ and \ZNA\ event classes. Nevertheless, within the uncertainties, both 
follow a common trend as a function of $\avg{\dNdeta}|_{|\eta|<0.5}$.

So far it has been seen that the assumption of an unmodified jet shape in different multiplicity classes in \pPb\ collisions 
resulted in the emergence of almost identical ridge-like excess structures on the near side and away side, most pronounced in 
high-multiplicity events. 
An alternative approach is to start with the assumption that there are identical ridge structures on the 
near side and away side, and to study whether this assumption leaves any room for multiplicity dependent modifications of the 
jet shape, in particular on the away side. 
To this end, a symmetric ridge structure is subtracted on the near side and away side from the $\Dphi$ projection of the associated 
yield per trigger averaged over $|\Deta|<1.8$.
The near-side ridge structure is determined in the same event class within $1.2<|\Deta|<1.8$, while the ridge on the away side 
is constructed by mirroring this near-side structure at $\Dphi=\pi/2$.
The ridge-subtracted results in the interval $2<\ptt<$~\unit[4]{\gevc} and $1<\pta<$~\unit[2]{\gevc} for the 0--20\%, 20--40\% and 40--60\% 
event classes are shown in \Fig{fig:1dcorr_ridgesubtracted} compared to the unsubtracted 60--100\% event class and to \pp\ collisions.
The remaining yields in all event classes are in agreement with each other and with \pp\ collisions, indicating that the observed 
correlations are indeed consistent with a symmetric ridge and with no further significant modification of the jet structure at midrapidity 
in high-multiplicity \pPb\ collisions at the LHC, in contrast to what was seen at forward rapidity in $\snn=$~\unit[0.2]{TeV} \dAu\ 
collisions at RHIC~\cite{Braidot:2010zh}.

\section{Summary}
\label{sec:summary}
Results from angular correlations between charged trigger and associated particles in \pPb\ collisions at $\snn=$~\unit[5.02]{TeV} 
are presented for various transverse momentum ranges within $0.5<\pta<\ptt<$~\unit[4]{\gevc}.
Associated yields per trigger particle are measured over two units of pseudorapidity and full 
azimuthal angle in different multiplicity classes.
The yields projected onto $\Dphi$ increase with event multiplicity 
and rise to values higher than those observed in \pp\ collisions at 
$\s =$ 2.76 and \unit[7]{TeV}.
The difference between the yields per trigger particle in high-multiplicity and low-multiplicity 
events exhibits two nearly identical, long-range (up to $|\Deta| \sim 2$) ridge-like excess structures 
on the near-side ($\Dphi \approx 0$) and away-side ($\Dphi \approx \pi$) as quantified by their yields and widths.
The excess on the near side at high event multiplicity is qualitatively similar to that recently reported by the CMS 
collaboration in $2<|\Deta|<4$~\cite{CMS:2012qk}. 
The excess on the away side with respect to the usual away-side structure due to back-to-back jets and momentum conservation
is reported here for the first time, and confirmed by a similar study 
from the ATLAS collaboration~\cite{atlasridge} that appeared after publication of this paper.
The event multiplicity and $\pt$ dependences of the near-side and away-side ridge yields are in good agreement, and
their widths show no significant dependence on multiplicity or $\pt$.
The observation of a nearly identical near-side and away-side ridge-like structure is consistent with Colour Glass Condensate model 
calculations~\cite{Dusling:2012wy}. 
At the same time, 
the extracted $v_2$ and $v_3$ coefficients are in qualitative agreement with a hydrodynamical model calculation~\cite{Bozek:2012gr}.
Further theoretical investigation is needed for a detailed understanding of the origin of these long-range 
correlation structures.
After subtracting the near-side ridge from the near side and away side symmetrically, the correlation shape
in $\Dphi$ becomes independent of multiplicity and similar to those of \pp\ collisions at \unit[7]{TeV}. 
There is no evidence in the present data for further significant structures in two-particle correlations at midrapidity 
in \pPb\ collisions at $\snn = \unit[5.02]{TeV}$.

\ifpreprint
\iffull
\newenvironment{acknowledgement}{\relax}{\relax}
\begin{acknowledgement}
\section*{Acknowledgements}
The ALICE collaboration would like to thank all its engineers and technicians for their invaluable contributions to the construction of the experiment and the CERN accelerator teams for the outstanding performance of the LHC complex.
\\
The ALICE collaboration acknowledges the following funding agencies for their support in building and
running the ALICE detector:
 \\
State Committee of Science, Calouste Gulbenkian Foundation from
Lisbon and Swiss Fonds Kidagan, Armenia;
 \\
Conselho Nacional de Desenvolvimento Cient\'{\i}fico e Tecnol\'{o}gico (CNPq), Financiadora de Estudos e Projetos (FINEP),
Funda\c{c}\~{a}o de Amparo \`{a} Pesquisa do Estado de S\~{a}o Paulo (FAPESP);
 \\
National Natural Science Foundation of China (NSFC), the Chinese Ministry of Education (CMOE)
and the Ministry of Science and Technology of China (MSTC);
 \\
Ministry of Education and Youth of the Czech Republic;
 \\
Danish Natural Science Research Council, the Carlsberg Foundation and the Danish National Research Foundation;
 \\
The European Research Council under the European Community's Seventh Framework Programme;
 \\
Helsinki Institute of Physics and the Academy of Finland;
 \\
French CNRS-IN2P3, the `Region Pays de Loire', `Region Alsace', `Region Auvergne' and CEA, France;
 \\
German BMBF and the Helmholtz Association;
\\
General Secretariat for Research and Technology, Ministry of
Development, Greece;
\\
Hungarian OTKA and National Office for Research and Technology (NKTH);
 \\
Department of Atomic Energy and Department of Science and Technology of the Government of India;
 \\
Istituto Nazionale di Fisica Nucleare (INFN) and Centro Fermi -
Museo Storico della Fisica e Centro Studi e Ricerche "Enrico
Fermi", Italy;
 \\
MEXT Grant-in-Aid for Specially Promoted Research, Ja\-pan;
 \\
Joint Institute for Nuclear Research, Dubna;
 \\
National Research Foundation of Korea (NRF);
 \\
CONACYT, DGAPA, M\'{e}xico, ALFA-EC and the HELEN Program (High-Energy physics Latin-American--European Network);
 \\
Stichting voor Fundamenteel Onderzoek der Materie (FOM) and the Nederlandse Organisatie voor Wetenschappelijk Onderzoek (NWO), Netherlands;
 \\
Research Council of Norway (NFR);
 \\
Polish Ministry of Science and Higher Education;
 \\
National Authority for Scientific Research - NASR (Autoritatea Na\c{t}ional\u{a} pentru Cercetare \c{S}tiin\c{t}ific\u{a} - ANCS);
 \\
Ministry of Education and Science of Russian Federation,
International Science and Technology Center, Russian Academy of
Sciences, Russian Federal Agency of Atomic Energy, Russian Federal
Agency for Science and Innovations and CERN-INTAS;
 \\
Ministry of Education of Slovakia;
 \\
Department of Science and Technology, South Africa;
 \\
CIEMAT, EELA, Ministerio de Educaci\'{o}n y Ciencia of Spain, Xunta de Galicia (Conseller\'{\i}a de Educaci\'{o}n),
CEA\-DEN, Cubaenerg\'{\i}a, Cuba, and IAEA (International Atomic Energy Agency);
 \\
Swedish Research Council (VR) and Knut $\&$ Alice Wallenberg
Foundation (KAW);
 \\
Ukraine Ministry of Education and Science;
 \\
United Kingdom Science and Technology Facilities Council (STFC);
 \\
The United States Department of Energy, the United States National
Science Foundation, the State of Texas, and the State of Ohio.
\end{acknowledgement}
\ifbibtex
\bibliographystyle{utphys}
\bibliography{biblio}{}
\else
\providecommand{\href}[2]{#2}\begingroup\raggedright\endgroup

\fi
\newpage
\appendix
\section{The ALICE Collaboration}
\label{app:collab}

\begingroup
\small
\begin{flushleft}
B.~Abelev\Irefn{org1234}\And
J.~Adam\Irefn{org1274}\And
D.~Adamov\'{a}\Irefn{org1283}\And
A.M.~Adare\Irefn{org1260}\And
M.M.~Aggarwal\Irefn{org1157}\And
G.~Aglieri~Rinella\Irefn{org1192}\And
M.~Agnello\Irefn{org1313}\textsuperscript{,}\Irefn{org1017688}\And
A.G.~Agocs\Irefn{org1143}\And
A.~Agostinelli\Irefn{org1132}\And
Z.~Ahammed\Irefn{org1225}
% ...
\renewcommand\labelenumi{\textsuperscript{\theenumi}~}
\section*{Affiliation notes}
\renewcommand\theenumi{\roman{enumi}}
\begin{Authlist}
\item \Adef{0}Deceased
\item \Adef{M.V.Lomonosov Moscow State University, D.V.Skobeltsyn Institute of Nuclear Physics, Moscow, Russia}Also at: M.V.Lomonosov Moscow State University, D.V.Skobeltsyn Institute of Nuclear Physics, Moscow, Russia
\item \Adef{University of Belgrade, Faculty of Physics and Vinca Institute of Nuclear Sciences, Belgrade, Serbia}Also at: University of Belgrade, Faculty of Physics and Vinca Institute of Nuclear Sciences, Belgrade, Serbia
\item \Adef{Institute of Theoretical Physics, University of Wroclaw, Wroclaw, Poland}Also at: Institute of Theoretical Physics, University of Wroclaw, Wroclaw, Poland
\end{Authlist}
\section*{Collaboration Institutes}
\renewcommand\theenumi{\arabic{enumi}~}
\begin{Authlist}
\item \Idef{org1332}A. I. Alikhanyan National Science Laboratory (Yerevan Physics Institute) Foundation, Yerevan, Armenia
\item \Idef{org1279}Benem\'{e}rita Universidad Aut\'{o}noma de Puebla, Puebla, Mexico
\item \Idef{org1220}Bogolyubov Institute for Theoretical Physics, Kiev, Ukraine
\item \Idef{org20959}Bose Institute, Department of Physics and Centre for Astroparticle Physics and Space Science (CAPSS), Kolkata, India
\item \Idef{org1262}Budker Institute for Nuclear Physics, Novosibirsk, Russia
\item \Idef{org1292}California Polytechnic State University, San Luis Obispo, California, United States
\item \Idef{org1329}Central China Normal University, Wuhan, China
\item \Idef{org14939}Centre de Calcul de l'IN2P3, Villeurbanne, France
\item \Idef{org1197}Centro de Aplicaciones Tecnol\'{o}gicas y Desarrollo Nuclear (CEADEN), Havana, Cuba
\item \Idef{org1242}Centro de Investigaciones Energ\'{e}ticas Medioambientales y Tecnol\'{o}gicas (CIEMAT), Madrid, Spain
\item \Idef{org1244}Centro de Investigaci\'{o}n y de Estudios Avanzados (CINVESTAV), Mexico City and M\'{e}rida, Mexico
\item \Idef{org1335}Centro Fermi - Museo Storico della Fisica e Centro Studi e Ricerche ``Enrico Fermi'', Rome, Italy
\item \Idef{org17347}Chicago State University, Chicago, United States
\item \Idef{org1288}Commissariat \`{a} l'Energie Atomique, IRFU, Saclay, France
\item \Idef{org15782}COMSATS Institute of Information Technology (CIIT), Islamabad, Pakistan
\item \Idef{org1294}Departamento de F\'{\i}sica de Part\'{\i}culas and IGFAE, Universidad de Santiago de Compostela, Santiago de Compostela, Spain
\item \Idef{org1106}Department of Physics Aligarh Muslim University, Aligarh, India
\item \Idef{org1121}Department of Physics and Technology, University of Bergen, Bergen, Norway
\item \Idef{org1162}Department of Physics, Ohio State University, Columbus, Ohio, United States
\item \Idef{org1300}Department of Physics, Sejong University, Seoul, South Korea
\item \Idef{org1268}Department of Physics, University of Oslo, Oslo, Norway
\item \Idef{org1312}Dipartimento di Fisica dell'Universit\`{a} and Sezione INFN, Turin, Italy
\item \Idef{org1145}Dipartimento di Fisica dell'Universit\`{a} and Sezione INFN, Cagliari, Italy
\item \Idef{org1315}Dipartimento di Fisica dell'Universit\`{a} and Sezione INFN, Trieste, Italy
\item \Idef{org1285}Dipartimento di Fisica dell'Universit\`{a} `La Sapienza' and Sezione INFN, Rome, Italy
\item \Idef{org1154}Dipartimento di Fisica e Astronomia dell'Universit\`{a} and Sezione INFN, Catania, Italy
\item \Idef{org1132}Dipartimento di Fisica e Astronomia dell'Universit\`{a} and Sezione INFN, Bologna, Italy
\item \Idef{org1270}Dipartimento di Fisica e Astronomia dell'Universit\`{a} and Sezione INFN, Padova, Italy
\item \Idef{org1290}Dipartimento di Fisica `E.R.~Caianiello' dell'Universit\`{a} and Gruppo Collegato INFN, Salerno, Italy
\item \Idef{org1103}Dipartimento di Scienze e Innovazione Tecnologica dell'Universit\`{a} del Piemonte Orientale and Gruppo Collegato INFN, Alessandria, Italy
\item \Idef{org1114}Dipartimento Interateneo di Fisica `M.~Merlin' and Sezione INFN, Bari, Italy
\item \Idef{org1237}Division of Experimental High Energy Physics, University of Lund, Lund, Sweden
\item \Idef{org1192}European Organization for Nuclear Research (CERN), Geneva, Switzerland
\item \Idef{org1227}Fachhochschule K\"{o}ln, K\"{o}ln, Germany
\item \Idef{org1122}Faculty of Engineering, Bergen University College, Bergen, Norway
\item \Idef{org1136}Faculty of Mathematics, Physics and Informatics, Comenius University, Bratislava, Slovakia
\item \Idef{org1274}Faculty of Nuclear Sciences and Physical Engineering, Czech Technical University in Prague, Prague, Czech Republic
\item \Idef{org1229}Faculty of Science, P.J.~\v{S}af\'{a}rik University, Ko\v{s}ice, Slovakia
\item \Idef{org1184}Frankfurt Institute for Advanced Studies, Johann Wolfgang Goethe-Universit\"{a}t Frankfurt, Frankfurt, Germany
\item \Idef{org1215}Gangneung-Wonju National University, Gangneung, South Korea
\item \Idef{org20958}Gauhati University, Department of Physics, Guwahati, India
\item \Idef{org1212}Helsinki Institute of Physics (HIP) and University of Jyv\"{a}skyl\"{a}, Jyv\"{a}skyl\"{a}, Finland
\item \Idef{org1203}Hiroshima University, Hiroshima, Japan
\item \Idef{org1254}Indian Institute of Technology Bombay (IIT), Mumbai, India
\item \Idef{org36378}Indian Institute of Technology Indore, Indore, India (IITI)
\item \Idef{org1266}Institut de Physique Nucl\'{e}aire d'Orsay (IPNO), Universit\'{e} Paris-Sud, CNRS-IN2P3, Orsay, France
\item \Idef{org1277}Institute for High Energy Physics, Protvino, Russia
\item \Idef{org1249}Institute for Nuclear Research, Academy of Sciences, Moscow, Russia
\item \Idef{org1320}Nikhef, National Institute for Subatomic Physics and Institute for Subatomic Physics of Utrecht University, Utrecht, Netherlands
\item \Idef{org1250}Institute for Theoretical and Experimental Physics, Moscow, Russia
\item \Idef{org1230}Institute of Experimental Physics, Slovak Academy of Sciences, Ko\v{s}ice, Slovakia
\item \Idef{org1127}Institute of Physics, Bhubaneswar, India
\item \Idef{org1275}Institute of Physics, Academy of Sciences of the Czech Republic, Prague, Czech Republic
\item \Idef{org1139}Institute of Space Sciences (ISS), Bucharest, Romania
\item \Idef{org27399}Institut f\"{u}r Informatik, Johann Wolfgang Goethe-Universit\"{a}t Frankfurt, Frankfurt, Germany
\item \Idef{org1185}Institut f\"{u}r Kernphysik, Johann Wolfgang Goethe-Universit\"{a}t Frankfurt, Frankfurt, Germany
\item \Idef{org1177}Institut f\"{u}r Kernphysik, Technische Universit\"{a}t Darmstadt, Darmstadt, Germany
\item \Idef{org1256}Institut f\"{u}r Kernphysik, Westf\"{a}lische Wilhelms-Universit\"{a}t M\"{u}nster, M\"{u}nster, Germany
\item \Idef{org1246}Instituto de Ciencias Nucleares, Universidad Nacional Aut\'{o}noma de M\'{e}xico, Mexico City, Mexico
\item \Idef{org1247}Instituto de F\'{\i}sica, Universidad Nacional Aut\'{o}noma de M\'{e}xico, Mexico City, Mexico
\item \Idef{org1308}Institut Pluridisciplinaire Hubert Curien (IPHC), Universit\'{e} de Strasbourg, CNRS-IN2P3, Strasbourg, France
\item \Idef{org1182}Joint Institute for Nuclear Research (JINR), Dubna, Russia
\item \Idef{org1199}Kirchhoff-Institut f\"{u}r Physik, Ruprecht-Karls-Universit\"{a}t Heidelberg, Heidelberg, Germany
\item \Idef{org20954}Korea Institute of Science and Technology Information, Daejeon, South Korea
\item \Idef{org1017642}KTO Karatay University, Konya, Turkey
\item \Idef{org1160}Laboratoire de Physique Corpusculaire (LPC), Clermont Universit\'{e}, Universit\'{e} Blaise Pascal, CNRS--IN2P3, Clermont-Ferrand, France
\item \Idef{org1194}Laboratoire de Physique Subatomique et de Cosmologie (LPSC), Universit\'{e} Joseph Fourier, CNRS-IN2P3, Institut Polytechnique de Grenoble, Grenoble, France
\item \Idef{org1187}Laboratori Nazionali di Frascati, INFN, Frascati, Italy
\item \Idef{org1232}Laboratori Nazionali di Legnaro, INFN, Legnaro, Italy
\item \Idef{org1125}Lawrence Berkeley National Laboratory, Berkeley, California, United States
\item \Idef{org1234}Lawrence Livermore National Laboratory, Livermore, California, United States
\item \Idef{org1251}Moscow Engineering Physics Institute, Moscow, Russia
\item \Idef{org1322}National Centre for Nuclear Studies, Warsaw, Poland
\item \Idef{org1140}National Institute for Physics and Nuclear Engineering, Bucharest, Romania
\item \Idef{org1017626}National Institute of Science Education and Research, Bhubaneswar, India
\item \Idef{org1165}Niels Bohr Institute, University of Copenhagen, Copenhagen, Denmark
\item \Idef{org1109}Nikhef, National Institute for Subatomic Physics, Amsterdam, Netherlands
\item \Idef{org1283}Nuclear Physics Institute, Academy of Sciences of the Czech Republic, \v{R}e\v{z} u Prahy, Czech Republic
\item \Idef{org1264}Oak Ridge National Laboratory, Oak Ridge, Tennessee, United States
\item \Idef{org1189}Petersburg Nuclear Physics Institute, Gatchina, Russia
\item \Idef{org1170}Physics Department, Creighton University, Omaha, Nebraska, United States
\item \Idef{org1157}Physics Department, Panjab University, Chandigarh, India
\item \Idef{org1112}Physics Department, University of Athens, Athens, Greece
\item \Idef{org1152}Physics Department, University of Cape Town and  iThemba LABS, National Research Foundation, Somerset West, South Africa
\item \Idef{org1209}Physics Department, University of Jammu, Jammu, India
\item \Idef{org1207}Physics Department, University of Rajasthan, Jaipur, India
\item \Idef{org1200}Physikalisches Institut, Ruprecht-Karls-Universit\"{a}t Heidelberg, Heidelberg, Germany
\item \Idef{org1017688}Politecnico di Torino, Turin, Italy
\item \Idef{org1325}Purdue University, West Lafayette, Indiana, United States
\item \Idef{org1281}Pusan National University, Pusan, South Korea
\item \Idef{org1176}Research Division and ExtreMe Matter Institute EMMI, GSI Helmholtzzentrum f\"ur Schwerionenforschung, Darmstadt, Germany
\item \Idef{org1334}Rudjer Bo\v{s}kovi\'{c} Institute, Zagreb, Croatia
\item \Idef{org1298}Russian Federal Nuclear Center (VNIIEF), Sarov, Russia
\item \Idef{org1252}Russian Research Centre Kurchatov Institute, Moscow, Russia
\item \Idef{org1224}Saha Institute of Nuclear Physics, Kolkata, India
\item \Idef{org1130}School of Physics and Astronomy, University of Birmingham, Birmingham, United Kingdom
\item \Idef{org1338}Secci\'{o}n F\'{\i}sica, Departamento de Ciencias, Pontificia Universidad Cat\'{o}lica del Per\'{u}, Lima, Peru
\item \Idef{org1133}Sezione INFN, Bologna, Italy
\item \Idef{org1271}Sezione INFN, Padova, Italy
\item \Idef{org1286}Sezione INFN, Rome, Italy
\item \Idef{org1146}Sezione INFN, Cagliari, Italy
\item \Idef{org1313}Sezione INFN, Turin, Italy
\item \Idef{org1316}Sezione INFN, Trieste, Italy
\item \Idef{org1115}Sezione INFN, Bari, Italy
\item \Idef{org1155}Sezione INFN, Catania, Italy
\item \Idef{org36377}Nuclear Physics Group, STFC Daresbury Laboratory, Daresbury, United Kingdom
\item \Idef{org1258}SUBATECH, Ecole des Mines de Nantes, Universit\'{e} de Nantes, CNRS-IN2P3, Nantes, France
\item \Idef{org35706}Suranaree University of Technology, Nakhon Ratchasima, Thailand
\item \Idef{org1304}Technical University of Split FESB, Split, Croatia
\item \Idef{org1168}The Henryk Niewodniczanski Institute of Nuclear Physics, Polish Academy of Sciences, Cracow, Poland
\item \Idef{org17361}The University of Texas at Austin, Physics Department, Austin, TX, United States
\item \Idef{org1173}Universidad Aut\'{o}noma de Sinaloa, Culiac\'{a}n, Mexico
\item \Idef{org1296}Universidade de S\~{a}o Paulo (USP), S\~{a}o Paulo, Brazil
\item \Idef{org1149}Universidade Estadual de Campinas (UNICAMP), Campinas, Brazil
\item \Idef{org1239}Universit\'{e} de Lyon, Universit\'{e} Lyon 1, CNRS/IN2P3, IPN-Lyon, Villeurbanne, France
\item \Idef{org1205}University of Houston, Houston, Texas, United States
\item \Idef{org20371}University of Technology and Austrian Academy of Sciences, Vienna, Austria
\item \Idef{org1222}University of Tennessee, Knoxville, Tennessee, United States
\item \Idef{org1310}University of Tokyo, Tokyo, Japan
\item \Idef{org1318}University of Tsukuba, Tsukuba, Japan
\item \Idef{org21360}Eberhard Karls Universit\"{a}t T\"{u}bingen, T\"{u}bingen, Germany
\item \Idef{org1225}Variable Energy Cyclotron Centre, Kolkata, India
\item \Idef{org1306}V.~Fock Institute for Physics, St. Petersburg State University, St. Petersburg, Russia
\item \Idef{org1323}Warsaw University of Technology, Warsaw, Poland
\item \Idef{org1179}Wayne State University, Detroit, Michigan, United States
\item \Idef{org1143}Wigner Research Centre for Physics, Hungarian Academy of Sciences, Budapest, Hungary
\item \Idef{org1260}Yale University, New Haven, Connecticut, United States
\item \Idef{org15649}Yildiz Technical University, Istanbul, Turkey
\item \Idef{org1301}Yonsei University, Seoul, South Korea
\item \Idef{org1327}Zentrum f\"{u}r Technologietransfer und Telekommunikation (ZTT), Fachhochschule Worms, Worms, Germany
\end{Authlist}
\endgroup

\else
\ifbibtex
\bibliographystyle{utphys}
\bibliography{biblio}{}
\else

\fi
\fi
\else
\iffull
\vspace{0.5cm}

\input{refpaper.tex}
\else
\ifbibtex
\bibliographystyle{model1-num-names}
\bibliography{biblio}{}
\else
\input{refpaper.tex}
\fi
\fi
\fi
\end{document}